\documentclass{article}
\usepackage{spconf,amsmath,graphicx}
\usepackage{makecell}
\usepackage{multirow}
\usepackage{xcolor}
\usepackage{amsfonts}
\usepackage{booktabs}
\usepackage{verbatim}

\usepackage{eso-pic}
\AddToShipoutPicture{%
  \AtPageUpperLeft{%
    \raisebox{-1cm}[0pt][0pt]{\makebox[\paperwidth]{\centering
\scriptsize This work has been submitted to the IEEE for possible publication. Copyright may be transferred without notice, after which this version may no longer be accessible.
}}%
  }%
}

\title{Teaching Audio Models to Reason: A Unified Framework for Source- and Layer-wise Distillation}

\name{ Runyan Yang$^{1,2,\dagger}$, Yuke Si$^{1,2,\dagger}$, Yingying Gao$^{1,2,\dagger}$, Junlan Feng$^{1,2}$, Chao Deng$^{1,2}$, Shilei Zhang$^{1,2,*}$\\ \thanks{† Equal contribution}\thanks{* Corresponding author}}
\address{ $^1$Jiutian Artificial Intelligence Research Institute, China Mobile, Beijing, China\\ 
$^2$The State Key Laboratory of Multimedia Information Processing, Peking University, Beijing, China
}

\begin{document}

%
\maketitle
\begin{abstract}
While large audio language models excel at tasks like ASR and emotion recognition, they still struggle with complex reasoning due to the modality gap between audio and text as well as the lack of structured intermediate supervision. To address this, we propose a unified knowledge distillation framework to transfer reasoning capabilities from a high-capacity textual teacher model to a student audio models while preserving its acoustic competence. Our method introduces two key dimensions: source-wise distillation, which leverages both textual and acoustic teachers to provide complementary modality-specific supervision; and layer-wise distillation, which aligns teacher signals with appropriate student layers to improve transfer efficiency. This dual-dimensional strategy enables fine-grained control over the distillation process, effectively bridging the gap between symbolic reasoning and speech representations. Experimental results show significant improvements in audio reasoning performance, demonstrating the effectiveness of our framework as a reasoning transfer solution for audio modeling.

\end{abstract}
\section{Introduction}
\label{sec:intro}
Recent advances in large audio language models (LALMs) have improved performance on tasks such as automatic speech recognition, speech translation, and emotion recognition~\cite{tang2023salmonn, deshmukh2023pengi, yang2024polyspeech}. However, their ability to perform complex reasoning over spoken content remains limited. Compared with text-based large language models, audio models face difficulties in multi-step reasoning due to the modality gap between audio and text as well as the lack of structured intermediate supervision during training.

To overcome the reasoning limitations of audio models, recent studies have explored large audio reasoning models (LARMs) that integrate structured prompting, chain-of-thought supervision, or reward shaping into audio models, enabling audio models to emulate step-wise reasoning similar to LLMs~\cite{ghosh2024gama,ma2025audio,li2025reinforcement}. While these approaches improve performance on complex auditory reasoning tasks, they typically require large-scale instruction tuning and substantial computational resources, limiting their practicality and scalability in real-world applications.

These challenges call for a more efficient and scalable solution to endow audio models with reasoning abilities. Knowledge distillation (KD) provides a natural solution by transferring skills from high-capacity teacher models to student models~\cite{jiang2023lion,gu2023minillm,ko2024distillm}. While KD has proven effective in textual domains, its use for structured reasoning in audio models remains underexplored. Moreover, conventional KD techniques assume fixed teacher sources and static supervision layer, which are not suited for the modality gap and representational hierarchy inherent in audio reasoning tasks.

In this work, we propose a unified and fine-grained distillation framework to teach audio models to reason by decoupling the supervision process into two dimensions: \textit{source-wise}, and \textit{layer-wise} distillation. 
\textit{Source-wise distillation} considers the origin and modality of the teacher model. The textual teacher offers strong capabilities in symbolic reasoning and commonsense inference, while the acoustic teacher 
provides modality-consistent supervision grounded in audio representation. We explore two source selection strategies. The first strategy employs only a textual teacher and avoids the input modality mismatch by aligning textual audio descriptions with raw audio. The second strategy leverages both audio and textual teachers, allowing the student to jointly learn from audio and text representations with complementary guidance.
\textit{Layer-wise distillation} addresses the architectural alignment between teacher and student, enabling the student to absorb relevant information at the most effective depths. We analyze how teacher modality and reasoning depth interact to guide supervision placement.

Together, these two dimensions form a reasoning-aware distillation framework tailored for audio models. Our experiments show that modeling source-wise and layer-wise interactions leads to significant improvements in reasoning accuracy, offering new insights into transferring reasoning abilities from LLMs to LALMs.


\begin{figure}[t]
\centering
\includegraphics[width=9cm]{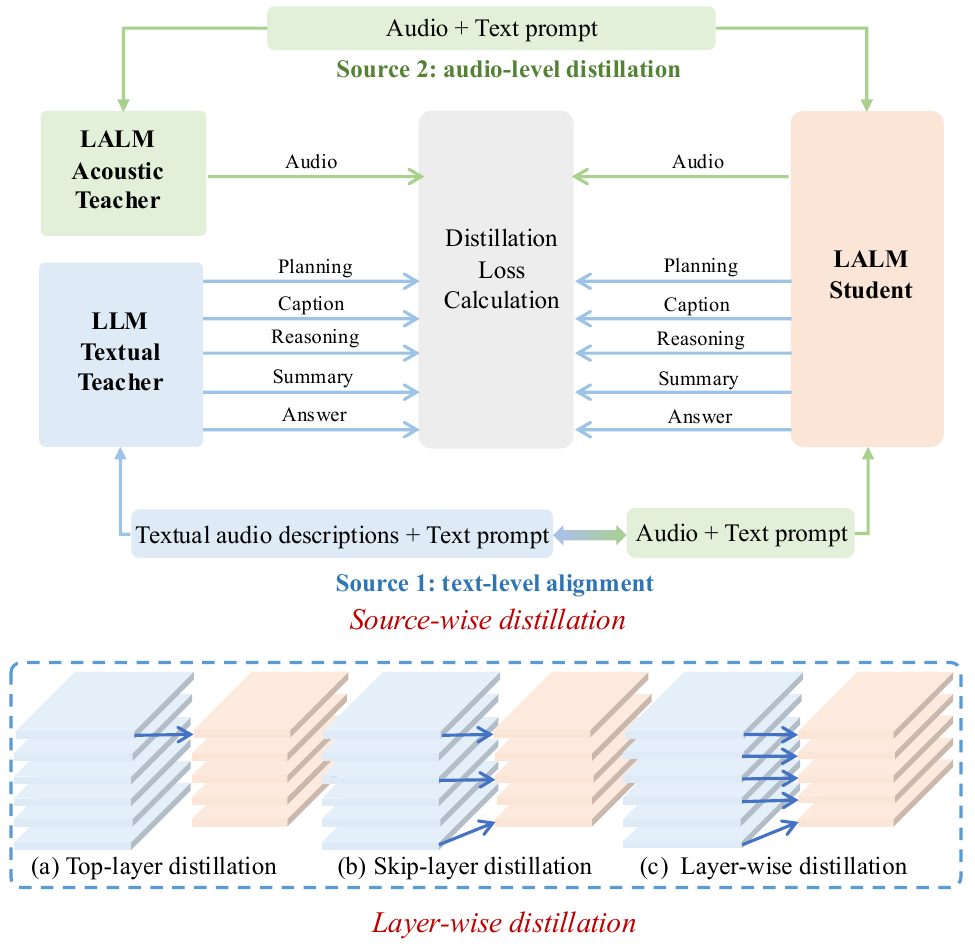}
\caption{Proposed teacher-student distillation framework}
\label{fig1}
\end{figure}

\section{Related Work}
\textbf{Large Audio Reasoning Models (LARMs).} LARMs are Large Audio Language Models (LALMs) that leverages the advanced reasoning capabilities of LLMs to understand complex queries with audio inputs. GAMA~\cite{ghosh2024gama} obtain complex reasoning abilities through instruction-tuning on LALM, by which the model is encouraged to analyze audio event according to the context such as other scene elements and world knowledge. CompA~\cite{ghosh2023compa} focuses on the compositional reasoning capacity of LALMs that attempts to understand the interrelationships, such as order of occurrence and attribute-binding, among acoustic events in an audio. Audio-CoT~\cite{ma2025audio} is the first exploration that integrates Chain-of-Thought (CoT) reasoning into LALMs to enhance their reasoning ability across auditory modalities. Audio-Reasoner~\cite{xie2025audio} is fine-tuned on Qwen2-Audio with structured CoT training. R1-AQA~\cite{li2025reinforcement} adopts reinforcement learning to improve the reasoning performance of the audio question answering (AQA) task. SARI~\cite{wen2025sari} compares explicit vs. implicit reasoning and structured vs. unstructured thinking process for LARMs. Audio Flamingo 3~\cite{goel2025audio} supports on-demand thinking and long audio understanding and reasoning. Audio-Thinker~\cite{wu2025audio} considers the question of when and how to think and incorporates multiple think rewards related to task complexity, the overall consistency and quality of the reasoning process, exhibiting State-of-the-Art performance on diverse benchmarks.

\textbf{Distillation of Large Language Models.} In LLMs scenarios, standard knowledge distillation objective becomes sub-optimal since the teacher model contains many more modes than student model. Therefore, more and more work is starting to consider the feedback from student model. Lion~\cite{jiang2023lion} is an adversarial distillation framework that incorporates the feedback of the student model and leverages the versatile role adaptability of LLMs, in which the teacher model is prompted to identify and generate “hard” instructions for student model to boost its proficiency iteratively. To prevent the student model from overestimating the low-probability regions of the teacher distribution due to the asymmetric nature of the Kullback-Leibler divergence (KLD), MiniLLM~\cite{gu2023minillm} adopts reverse KLD (RKL) to replace the forward KLD objective. Similarly, DISTILLM~\cite{ko2024distillm} introduces skew KLD (SKL), f-DISTILL~\cite{wen2023f} proposes Jenson–Shannon distillation (JSD), DISTILLM-2~\cite{ko2025distillm} integrates SKL and SRKL and achieves faster convergence and greater effectiveness. Besides the distillation loss, some work focuses on the distillation process in a white-box manner. Distilling step-by-step~\cite{hsieh2023distilling} adopts LLM to extract rationales as additional supervision for training small models within a multi-task framework. DDK~\cite{liu2024ddk} controls the composition of the distillation dataset according the performance differences between the teacher and student models.

\section{Method}\label{sec:method}

The proposed distillation framework is illustrated in Fig.~\ref{fig1}. Through \textit{Source-wise distillation}, the LALM student utilizes the knowledge of the LLM textual teacher and the LALM acoustic teacher together. Through \textit{Layer-wise distillation}, the teachers guide the student using information at various depths. We organize this section as follows: first, we describe the textualization of audio, which is the foundation for textual distillation; then, we illustrate layer-wise distillation in the context of textual distillation; finally, we introduce the acoustic distillation approach and the joint training objective.

\subsection{Textualization of audio}
\label{textualize}

A key challenge we face is that the textual teacher cannot directly process audio inputs. To bridge this modality gap, we design a textualization method that converts audio into textual descriptions. This enables the textual teacher to operate in its native modality while still providing reasoning supervision aligned with the audio.

To construct textual audio descriptions, we utilize the CoTA dataset~\cite{xie2025audio}, which was introduced to improve the reasoning ability of LALMs with structured CoT training. The CoTA dataset is denoted by 
\begin{equation}
\mathcal{D}_{\rm audio} = \left\{\left( x_i, q_i, r_i, a_i \right)\right\}_{i=1}^N,
\end{equation}
where $N$ is the dataset size. Each sample contains an audio input $x$, a textual question $q$, a four-stage reasoning trace $r=\{r^{(j)}\mid 1 \leq j \leq 4\}$ consisting of (1) \textit{planning}, (2) \textit{caption}, (3) \textit{reasoning}, and (4) \textit{summary}, as well as a final answer $a$. The reasoning task is to predict $r$ and $a$ given $x$ and $q$. We instruct an LALM to extract from the reasoning trace a concise audio description $d$, which captures audio content including essential information that supports subsequent reasoning. 
This process yields a textualized dataset: 
\begin{equation}
\mathcal{D}_{\rm text} = \left\{\left( d_i, q_i, r_i, a_i \right)\right\}_{i=1}^N.
\end{equation}
The prompt we use to instruct the LALM is presented below:

\begin{center}
\fcolorbox{black}{gray!10}{\parbox{0.95\linewidth}{\footnotesize You are an excellent audio analyst. Next, you will receive an audio and a question about this audio. You will also receive an reasoning trace, which involves some absolutely correct information about this audio. Your task is to analyze the audio content and generate a detailed textual description that includes all information from the audio relevant to the question-answering task, such that another model, which only processes text and does not have access to the original audio, can accurately answer the question based solely on your description. The audio description you provide should not be in conflict with the information from the given reasoning trace.

Your description may include the following aspects:

    1. What the speaker(s) said (verbatim or summarized);
    2. If there are multiple speakers, identify them and indicate the order of their speech;
    3. Speaking tone, emotion, and emphasis (if helpful for understanding the question);
    4. Key facts, background information, and reasoning cues mentioned in the audio;
    5. Significant pauses, hesitations, or emphasis in speech if relevant;
    6. Any background or environmental sounds that might be relevant (e.g., car sounds, music).

Do not add unrelated subjective interpretations or opinions—just objectively reconstruct everything in the audio that could assist in answering the question.

Below is the audio and its corresponding question and reasoning trace:

Here is the audio.

Here is the question: \textbf{**Question**}

Here is the reasoning trace: \textbf{**Reasoning trace**}

Please output a textual description of the audio that is suitable for answering the question:}}
\end{center}

\subsection{Layer-wise KD}
The conventional knowledge distillation method minimizes a divergence measure, e.g., Kullback-Leibler divergence (KLD) or Jensen-Shannon divergence (JSD), between the teacher’s and student’s predictive distributions at their top layers. Taking the textual distillation as an example, the objective for each output step $t$ is:
\vspace{-3pt}
\begin{equation}
\mathcal{L}_{{\rm top},t} = 
{\rm KD}\big( p_{\theta_T}(y_t \mid d, q, y_{<t}) \;\|\; p_{\theta_S}(y_t \mid x, q, y_{<t}) \big),
\end{equation}
where $y_t=\left\{ r,a \right\}_t$ is the token that the model predicts, and ${\rm KD}\left(\cdot\|\cdot\right)$ is the divergence measure.

Similarly to~\cite{sun2019patient,romero2014fitnets,zhang2024autocorrelation}, in our distillation framework, the knowledge of the teacher model is distilled not only to the student’s top-layer, but also to the student's each layer's representations. This layer-wise distillation allows the student to capture hierarchical feature representations learned by the teacher, leading to richer and more structured knowledge transfer. 

Since the number of layers in the textual teacher model may not be an integer multiple of that in the student model, it is not always feasible to align layers by simple skipping. Instead, we align them proportionally: for the $l_i^S$-th student layer, its corresponding teacher layer index $l_i^T$ is determined by
\vspace{-3pt}
\begin{equation}
l_i^T = \left\lfloor\frac{l_i^S-1}{L_{S}} \cdot L_{T} + 1\right\rfloor,
\end{equation}

where $L_{S}$ and $L_{T}$ are the numbers of layers of the student and teacher model, respectively. The layer-wise KD training objective for each layer at each output step is: 
\vspace{-3pt}
\begin{equation}
\mathcal{L}_{{\rm layer},i} =\mathrm{KD}\Big(W_i
h^{T}_{i,t} \;\Big\|\; h^{S}_{i,t}
\Big),
\end{equation}

where $h^{T}_{i,t} \in \mathbb{R}^{D_T}$ and $h^{S}_{i,t} \in \mathbb{R}^{D_S}$ are the hidden representations of the textual teacher model's $l_i^T$-th layer and the student model's $l_i^S$-th layer, respectively. $W_i \in \mathbb{R}^{D_T \times D_S}$ is a layer-specific learnable matrix, which maps $h^{T}_{i,t}$ to the same dimension as $h^{S}_{i,t}$. $\alpha_{\rm layer}$ is a scaling hyperparameter. The training objective for the whole output text token sequence is 

\vspace{-10pt}
\begin{equation}
\mathcal{L}_{\rm txt} =
\sum_{t \in \mathcal{T}_y}
\left(\mathcal{L}_{{\rm top},t} +
 \alpha_{\rm layer} \sum_{i=1}^{L_S}
 \mathcal{L}_{{\rm layer},i}\right),
\end{equation}
where $\mathcal{T}_y$ are a set of token indices that the model predicts.

We also propose an additional setting in which one layer is distilled every $k$ layers (1-in-$k$), as an intermediate approach between distilling all layers and distilling only the top layer. This approach is referred to as skip-layer distillation. The training objective is defined as follows:

\vspace{-10pt}
\begin{equation}
\mathcal{L}_{\rm txt,SL} =
\sum_{t \in \mathcal{T}_y}
\left(\mathcal{L}_{{\rm top},t} +
 \alpha_{\rm layer} \sum_{i=1}^{L_S / k}
 \mathcal{L}_{{\rm layer},ki}\right).
\end{equation}

\subsection{Acoustic KD}

We additionally perform representation distillation on the hidden states corresponding to the input audio tokens, in order to preserve the model’s ability to process acoustic representations and thus maintain fundamental acoustic capability. We refer to a frozen snapshot of the pre-trained LALM student taken before distillation, $S0$, as the acoustic teacher. As the LALMs does not yield logit outputs at the time steps corresponding to audio tokens, we only perform hidden-state distillation for acoustic KD. 
The acoustic distillation loss is
\begin{equation}
\mathcal{L}_{\rm ac} =
\sum_{t \in \mathcal{T}_{x}}
\sum_{i=1}^{L_S}
 \mathrm{KD}\Big(
h^{S0}_{i,t} \;\Big\|\; h^{S}_{i,t}
\Big),
\end{equation}
where $\mathcal{T}_{x}$ is the set of token positions corresponding to the input audio, and $h^{S0}_{i,t}$ denotes the hidden representation at the $l_i^S$-th layer of the acoustic teacher. 

\subsection{Joint Training Objective}
By combining above KD objectives and a supervised fine-tuning objective, we define the final joint training loss as:
\vspace{-3pt}
\begin{equation}
\mathcal{L}_{\rm joint} =
\mathcal{L}_{\rm txt} +
\alpha_{\rm ac} \mathcal{L}_{\rm ac} +
\alpha_{\rm SFT}  \mathcal{L}_{\rm SFT},
\end{equation}
where $\mathcal{L}_{\rm SFT}$ is the conventional cross-entropy loss used for supervised fine-tuning of the student model. $\alpha_{\rm ac}$ and  $\alpha_{\rm SFT}$ are weight coefficients (hyperparameters).

\section{Experiments}\label{sec:exp_res}

\subsection{Datasets}\label{sec:data}
For training, we utilize the CoTA dataset~\cite{xie2025audio}. 
The audio description introduced in Section~\ref{textualize} is generated using Qwen2.5-Omni-7B~\cite{xu2025qwen2} with greedy search. 
For evaluation, we mainly assess our method on an open-source audio question answering (AQA) benchmark MMAU (v05.15.25, test-mini subset)~\cite{sakshi2024mmau}, which covers multiple domains (sound, music, and speech), various reasoning / information extraction skills, and different difficulty levels. We also present results on  speech emotion recognition (SER) benchmark IEMOCAP (session 5)~\cite{busso2008iemocap} as supplementary reference.
No evaluation data are used in model training.


\subsection{Model Training Setup}\label{sec:expsetup}
In our knowledge distillation framework, we adopt Qwen2.5-Omni-7B thinker~\cite{xu2025qwen2} as the student model, initialized from its pre-trained parameters, and employ Qwen3-8B~\cite{qwen2025qwen3} as the textual teacher model. The Transformer layer numbers of the student and the textual teacher are 28 and 36, respectively.

We train the model for 3 epochs, setting the maximum learning rate to 1e-5. $\alpha_{\rm layer}$, $\alpha_{\rm ac}$, and $\alpha_{\rm SFT}$ are set to 0.05, 0.05, and 0.5, respectively. We adopt JSD as the KD divergence measure because it is symmetric and bounded, and it yields more stable training than KLD. Model training is performed on 8 NVIDIA A800 (80GB) GPUs.

\subsection{Model Inference Setup}\label{sec:infsetup}
During inference, we use the same generation parameters across all experimental settings: temperature = 0.6, top-k = 5, and top-p = 0.5.
For more precise and reliable evaluation for AQA, we standardize the final answer generated by the LALM to fit MMAU's evaluation script. We discard the generated reasoning trace in the evaluation.

\subsection{Results and analysis}

\begin{table}[]
\caption{Experimental results}\label{tab1}
\renewcommand{\arraystretch}{1.2}
\resizebox{1\linewidth}{!}{%
\begin{tabular}{lcc} 
\hline

\multirow{2}{*}{\makecell{\\[-0.9ex] Method}} & AQA Acc. (\%) & \multirow{2}{*}{\makecell{\\[-1ex] SER \\[-0.5ex] UA(\%)}} \\  
\cline{2-2}
& \makecell{ Sound / Music \\[-0.5ex]  Speech / Average} & \\ 
\hline 

Baseline & \makecell{ 74.47 / 66.47 \\[-0.5ex]  70.27 / 70.40} & \textbf{58.89} \\ 

SFT-only & \makecell{ 69.37 / 68.56 \\[-0.5ex]  71.47 / 69.80} & 51.93 \\ 

Top-layer txt KD + SFT & \makecell{ 70.57 / 66.47 \\[-0.4ex]  73.87 / 70.30} & 54.13 \\ 

Skip-layer txt KD (1-in-7) + SFT & \makecell{ 70.87 / 68.86 \\[-0.5ex]  72.37 / 70.70} & 53.37 \\ 

Layer-wise txt KD + SFT  & \makecell{ 70.87 / \textbf{70.96} \\[-0.5ex]  \textbf{75.68} / 72.50} & 49.65 \\ 

Layer-wise txt KD + ac KD + SFT & \makecell{ \textbf{75.38} / 70.36 \\[-0.4ex]  74.17 / \textbf{73.30} } & 56.03 \\ 

\hline       
\end{tabular}%
}
\vspace{-10pt}
\end{table}

Experimental results are illustrated in Table~\ref{tab1}. \textit{Baseline} refers to the results reproduced using the original Qwen2.5-Omni-7B model. We report accuracies for AQA and GID. For SER, we evaluate unweighted accuracy (UA), which averages accuracies over classes (happy, anger, sad, and neutral).

The results indicate that simple supervised fine-tuning (\textit{SFT-only}) does not yield consistent gains over the baseline. While SFT slightly improves AQA performance on speech questions, it degrades results on sound-related questions and SER, suggesting that naive SFT introduces catastrophic forgetting across heterogeneous speech tasks.

Incorporating knowledge distillation provides more stable improvements. \textit{Top-layer txt KD} surpasses \textit{SFT-only} on both AQA and SER, though its gains on AQA remain limited, highlighting the insufficiency of relying solely on the final representation. 
\textit{Layer-wise txt KD} further boosts AQA accuracy, reaching the best performance on speech-related questions (75.68\%), but at the cost of degraded SER. This suggests that fully distillation at all depths can overfit textual reasoning ability tasks while neglecting audio-related abilities.
As expected, \textit{Skip-layer txt KD (1-in-7)} achieves intermediate performance between top-layer KD and layer-wise KD.

Finally, combining \textit{Layer-wise txt KD} and \textit{ac KD} yields the overall best performance on AQA (average 73.30\%). Comparing to \textit{Layer-wise txt KD + SFT}, the incorporation of acoustic distillation brings substantial improvements on sound AQA (+4.51\%) and SER (+6.38\%), indicating that it helps maintain the model’s abilities to perceive and analyze low-level acoustic features. It is also observed that our LALMs trained using the CoTA dataset underperform the baseline in SER performance. This is because CoT reasoning may leverage semantic cues, which can occasionally misguide the model’s inference. 

\section{Conclusion}\label{sec:conclusion}
In this work, we propose a fine-grained distillation framework to equip audio models with reasoning abilities. 
Our approach introduces source-wise and layer-wise supervision to address the modality gap and architectural misalignment between teacher and student models.
By leveraging complementary strengths of textual and acoustic teachers and aligning their signals with appropriate student layers, our method enables more effective knowledge transfer.
Experiments demonstrate that the dual-dimensional strategy significantly improves reasoning performance, offering a new solution for   transferring reasoning capabilities from LLMs to LALMs.

\small
\bibliographystyle{IEEE}
\bibliography{mybib}

\end{document}